\documentclass{webofc}
\usepackage[varg]{txfonts}   % Web of Conferences font
\usepackage{hyperref}
\usepackage{url}
\newcommand{\Xmax}{$X_{\mathrm{max}}$}

\hypersetup{colorlinks=true,citecolor=blue,urlcolor=blue,linkcolor=blue}
\begin{document}
\title{Energy evolution of cosmic-ray mass and intensity measured by the Pierre Auger Observatory}
%
% subtitle is optionnal
%
%%%\subtitle{Do you have a subtitle?\\ If so, write it here}

%\author{\firstname{Vladim\'ir} \lastname{Novotn\'y}\inst{1}\fnsep\thanks{\email{novotnyv@ipnp.mff.cuni.cz}} for the Pierre Auger Collaboration\inst{2}\fnsep\thanks{\email{spokespersons@auger.org}}
%}
\author{\firstname{Vladim\'ir} \lastname{Novotn\'y}\inst{1}~for the Pierre Auger Collaboration\inst{2}\fnsep\thanks{\email{spokespersons@auger.org}}
}

\institute{Institute of Particle and Nuclear Physics, Faculty of Mathematics and Physics,
Charles University, V~Hole\v sovi\v ck\'ach 2, 180 00 Prague 8, Czech Republic
\and
Observatorio Pierre Auger, Av. San Mart\'in Norte 304, 5613 Malarg\"ue, Argentina\\
Full author list: \href{http://www.auger.org/archive/authors_2024_09.html}{http://www.auger.org/archive/authors\_2024\_09.html}
}

\abstract{The Pierre Auger Observatory has conducted measurements of the energy spectrum and mass composition of cosmic rays using different methods.
Utilizing both surface and fluorescence detectors (SD and FD), the Observatory provides unprecedented precision in understanding these particles.
While primarily designed to measure ultra-high energy cosmic rays, the FD’s high-elevation telescopes and the dense arrays of SD stations enable observations down to 6 PeV and 60 PeV, respectively.
To determine the depth of shower maximum, a critical parameter for identifying primary particle types, both direct longitudinal profile measurements from the FD and indirect signal analyses from the SD are employed.
An energy evolution of the mass of primary particles, as well as of the spectral index of the flux intensity, are observed and characterized by features described in the presented work.
The measurements benefit from the joint operation of the FD and SD, delivering a systematic uncertainty of $14\%$ in energy determination and an accumulated exposure reaching 80 000$\,\rm{km}^2$\,sr\,yr at the highest energies.
}
\maketitle
\section{Introduction}
\label{sec:intro}
The Pierre Auger Observatory \cite{PierreAuger:2015eyc} is the largest cosmic-ray observatory built to date, covering an area of $\sim$3000\,km$^2$.
Using its fluorescence and surface detectors (FD and SD), it measures the intensity of incoming cosmic rays and their mass composition through the induced extensive air showers (EAS), as described in Sections~\ref{sec:spectrum} and~\ref{sec:mass}, respectively.
The SD comprises arrays of water-Cherenkov stations, one with the spacing of 1500\,m, dedicated to ultra-high energies above 2\,EeV, the 750\,m array sensitive to showers above 100\,PeV, and the 433\,m array with full trigger efficiency above 60\,PeV.
The FD consists of 24 telescopes overlooking the 1500\,m array of the SD, and 3 high-elevation telescopes that investigate lower energies down to 6\,PeV.

\section{Energy spectrum}
\label{sec:spectrum}

The intensity of cosmic rays is measured separately using each of the SD arrays \cite{spec1500_PhysRevD.102.062005, infill_PierreAuger:2021hun, BrichettoOrquera:202340}, with the use of the hybrid method, and solely by the FD in Cherenkov regime \cite{Novotný:2021sA}.
As described in Refs.~\cite{spec1500_PhysRevD.102.062005, Novotný:2021sA}, the SD measurements are calibrated with FD-determined energies on a common subset of events.
This method keeps the huge statistics collected by the SD which operates constantly, while the more precise FD data, limited to clear moonless nights,
%that reduce its duty cycle to $\sim$14\,\%, 
deliver an energy-scale uncertainty of 14\,\% \cite{Dawson:2019pk}.

\begin{figure}[!h]
\centering
\includegraphics[width=66mm]{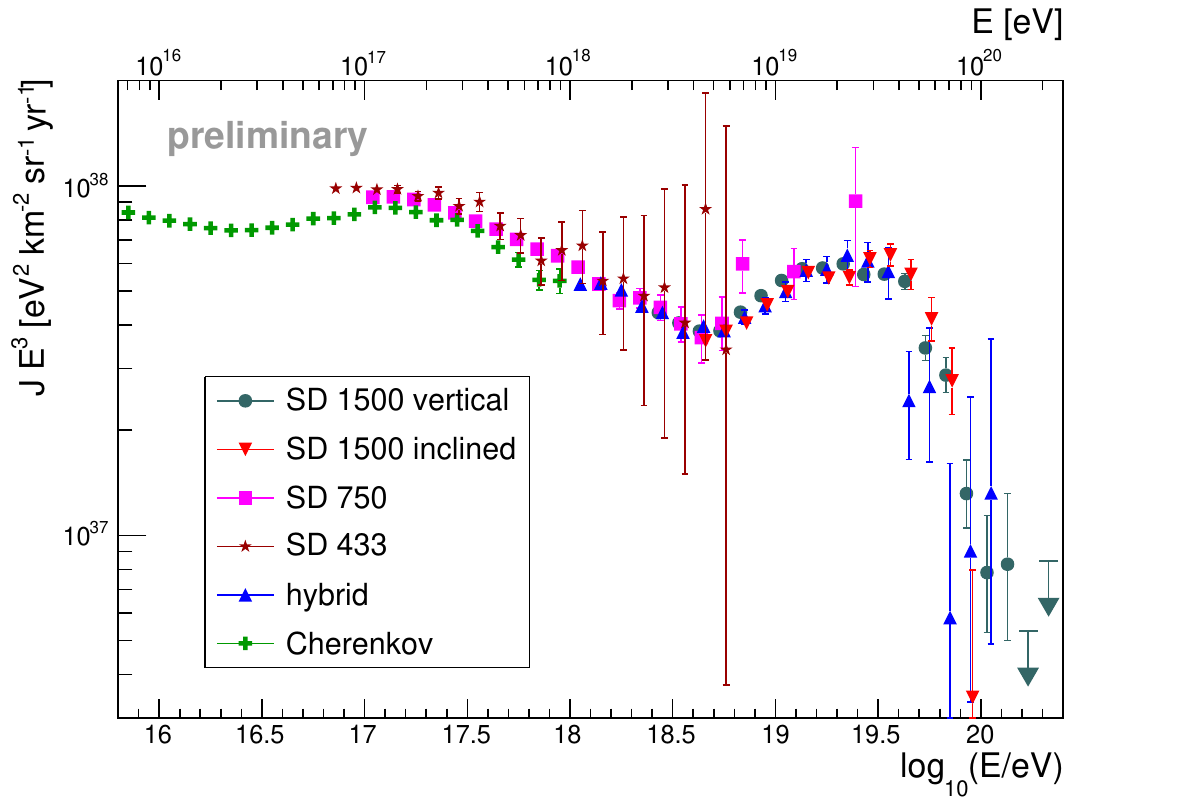}
\includegraphics[width=63mm]{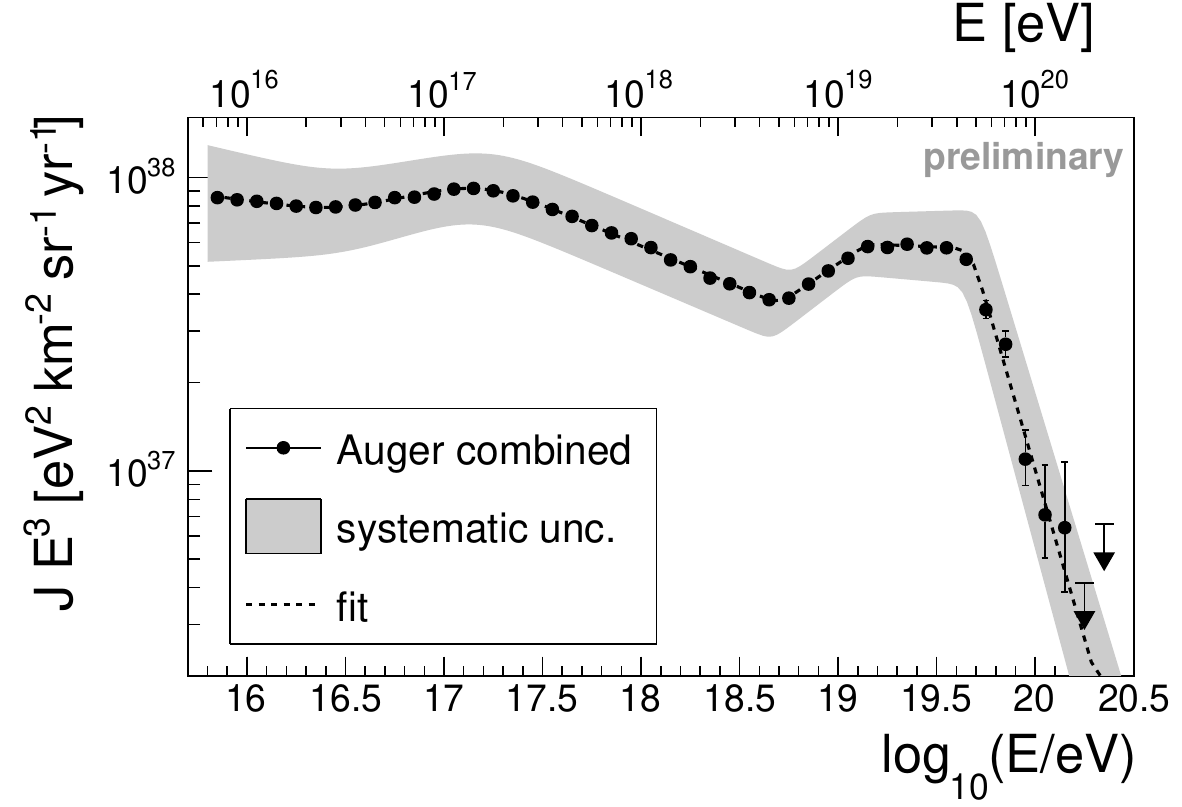}
\caption{
\label{fig:spectra}
The energy spectrum of cosmic rays derived from six data sets of the Pierre Auger Observatory (left) and the combined spectrum (right).
Data come from Refs.~\cite{Novotný:2021sA, BrichettoOrquera:202340}.
}
%\vspace{-4mm}
\end{figure}

Individual estimates of cosmic-ray intensity are shown in Fig.~\ref{fig:spectra}, together with the combined spectrum where the intensity is modeled as a sequence of power laws with parameters listed in Ref.~\cite{Novotný:2021sA}.

\section{Mass composition}
\label{sec:mass}

\begin{figure}[!h]
\centering
\includegraphics[width=60mm]{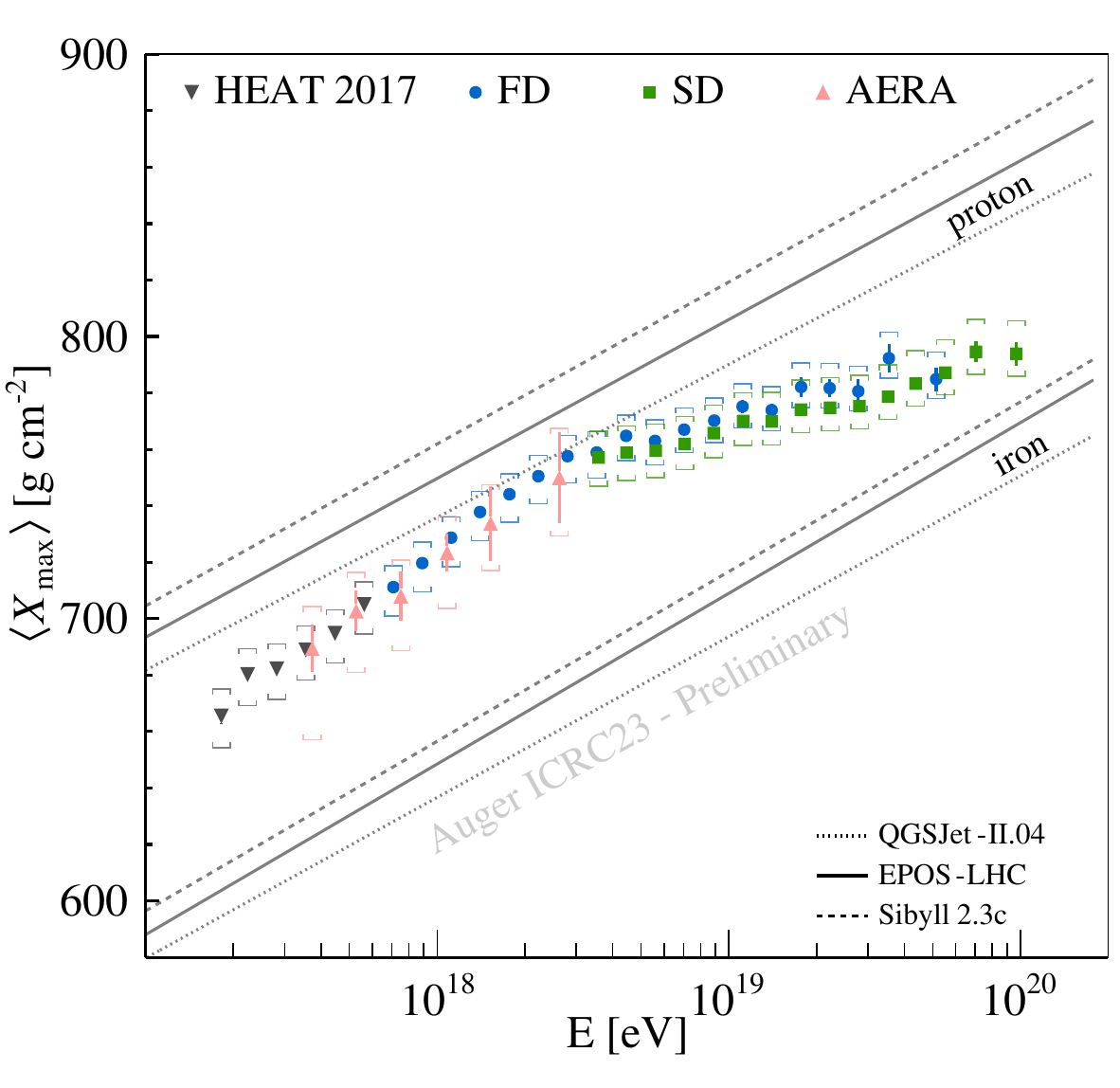}
\includegraphics[width=60mm]{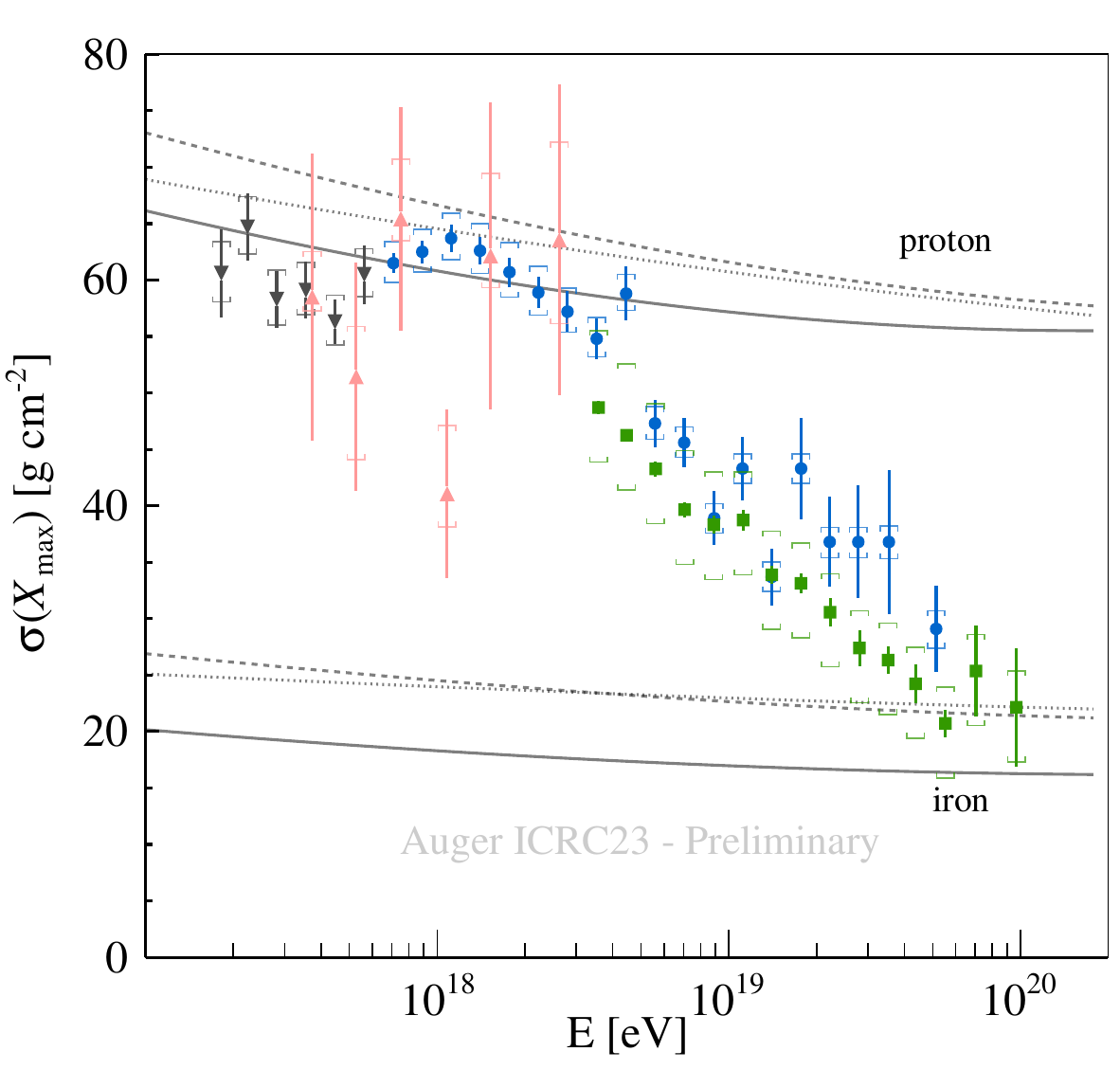}
\caption{
\label{fig:xmax}
Energy evolution of the average (left) and standard deviation (right) of \Xmax~distribution.
Data from horizontally-looking telescopes  are marked as FD \cite{xmaxfd_AbdulHalim:20239/} while high-elevation data as HEAT~2017 \cite{Bellido:2017Li}.
The estimate from deep learning is given as SD \cite{deep_AbdulHalim:2023C3}, and AERA shows
the radio dataset \cite{radio_PhysRevD.109.022002}.
Reproduced from Ref.~\cite{Mayotte:2023Nc}.
}
\vspace{-2mm}

\end{figure}

At the Pierre Auger Observatory, the mass composition of cosmic rays is determined from the depth of the EAS maximum, \Xmax.
This characteristic is accessible either directly or indirectly.
In the direct method, the longitudinal profile of the EAS is measured using the FD \cite{xmaxfd_AbdulHalim:20239/} or a radio array \cite{radio_PhysRevD.109.022002}.
One indirect method uses signal traces in water-Cherenkov stations of the SD processed by deep-learning algorithms \cite{deep_AbdulHalim:2023C3}.
The average and standard deviation of \Xmax~distribution obtained from both types of methods are shown in Fig.~\ref{fig:xmax} as functions of energy.
The \Xmax~scale reported by the SD measurement is calibrated with the FD data on a common subset, thus these are by construction consistent within systematic uncertainties.

\begin{figure}[!h]
\centering
\includegraphics[width=70mm]{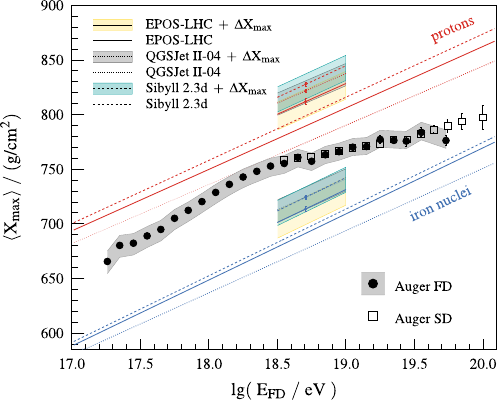}
\caption{
\label{fig:vicha}
Average \Xmax~measured at the Pierre Auger Observatory accompanied by predictions of unmodified (without bands) and modified (with bands) models of hadronic interactions.
The \Xmax\ modifications are needed to describe the FD and SD data together.
Bands correspond to systematic uncertainties.
Taken from Ref.~\cite{vicha_PhysRevD.109.102001}.
}
\vspace{-5mm}
\end{figure}

Moments as well as full \Xmax~distributions can be interpreted in terms of logarithm of the mass number, $\ln{A}$.
Particular models of hadronic interactions must be used for this purpose, such a study is performed in Ref.~\cite{Mayotte:2023Nc}.
However, recent results show that contemporary interaction models are unable to consistently describe both SD and FD data of the Pierre Auger Observatory together \cite{vicha_PhysRevD.109.102001}, which leads to modified predictions of \Xmax, shown in Fig.~\ref{fig:vicha}, and thus a changed $\ln{A}$ interpretation.

\section{Conclusions}
\label{sec:conclusins}

After 17 years of operation, the Pierre Auger Observatory has measured the energy spectrum of ultra-high energy cosmic rays as well as their mass composition.
Both observables evolve with energy, suggesting complex characteristics of the sources of primary cosmic rays.
The interactions of primaries with electromagnetic radiation during propagation also affect the spectrum and composition seen at Earth \cite{combined_PierreAuger:2023htc}. 

The features in the energy spectrum are
named the {\it low-energy ankle}, the {\it $2^{nd}$ knee}, the {\it ankle}, the {\it instep} and a steep suppression above 47~EeV.
The Peters' cycle \cite{Peters:1961mxb} seems to be a good framework to describe the mass composition which is dominated by protons around 1 EeV, by helium nuclei at about 10 EeV, and by the CNO group around 50 EeV and above \cite{xmaxfd_AbdulHalim:20239/}.
However, this interpretation of \Xmax~data is valid only in the context of contemporary models of hadronic interactions.
Because these models are unable to describe all available Auger data together, an even heavier composition could be expected when parameters of interaction models are adjusted.

\section*{Acknowledgements}
This work was co-funded by the European Union and supported by the Czech Ministry of Education, Youth and Sports (Project No. FORTE - CZ.02.01.01/00/22\_008/0004632).

%
% BibTeX or Biber users please use (the style is already called in the class, ensure that the "woc.bst" style is in your local directory)
\bibliography{spectrum_mass} % Replace "your_bib_file" with the actual name of your .bib file
%
% Non-BibTeX users please use
%
%\begin{thebibliography}{}
%
% and use \bibitem to create references.
%
%\bibitem{RefJ}
% Format for Journal Reference
%Journal Author, Article title. Journal \textbf{Volume}, page numbers (year). \url{https://doi.org/Article-DOI-number}
% Format for books
%\bibitem{RefB}
%Book Author, \textit{Book title} (Publisher, place, year) page numbers
% etc
%\end{thebibliography}

\end{document}